\documentclass{iau}

\usepackage{amsmath}
\usepackage{graphicx}
\usepackage{multirow}
\usepackage{float}

\begin{document}

\lefttitle{Wang et al. 2024, 2025}
\righttitle{GS$^3$ Hunter collaboration}

\jnlPage{1}{7}
\jnlDoiYr{2025}
\doival{10.1017/xxxxx}

\aopheadtitle{UniversAI: Exploring the Universe with Artificial Intelligence \\ Proceedings IAU Symposium No. 397, 2025}
\editors{Ioannis Liodakis, Vasilis Efthymiou  \&  Maria Dainotti, eds.}

\title{Towards Understanding the Milky Way's Matter Field and
Dynamical Accretion History based on AI-GS$^3$ Hunter}

\author{Hai-Feng Wang$^{1}$, Guan-Yu Wang$^{1}$, Giovanni Carraro$^1$, Yuan-Sen Ting$^2$, Thor Tepper-Garc\'ia$^3$, Joss Bland-Hawthorn$^3$, Jeffrey Carlin$^4$, Yang-Ping Luo$^5$}
\affiliation{$^1$Dipartimento di Fisica e Astronomia ``Galileo Galilei", Universit\'a degli Studi di Padova, Vicolo Osservatorio 3, I-35122, Padova, Italy}
\affiliation{$^2$Research School of Astronomy \& Astrophysics, Australian National University, Canberra, ACT 2611, Australia}
\affiliation{$^3$Sydney Institute for Astronomy, School of Physics, University of Sydney, NSW 2006, Australia}
\affiliation{$^4$Rubin Observatory/Legacy Survey of Space and Time (LSST), 950 N. Cherry Ave. Tucson, AZ, 85719, USA}
\affiliation{$^5$Department of Astronomy, China West Normal University, Nanchong, 637002, P.\,R.\,China}

\begin{abstract}

We present GS$^{3}$ Hunter (Galactic-Seismology Substructures and Streams Hunter), a novel deep-learning method that combines Siamese Neural Networks and K-means clustering to identify substructures and streams in stellar kinematic data. Applied to Gaia EDR3 and GALAH DR3, it recovers known groups (e.g., Thamnos, Helmi, GSE, Sequoia) and, with DESI dataset, reveals that GSE consists of four distinct components (GSH-GSH1 through GSE-GSH4), implying a multi-event accretion origin. Tests on LAMOST K-giants recover Sagittarius, Hercules–Aquila, and Virgo Overdensity, while also uncovering new substructures. Validation with FIRE simulations shows good agreement with previous results. GS$^{3}$ Hunter thus offers a powerful tool to understand the Milky Way’s halo assembly and tidal history.

\end{abstract}


\maketitle
\vspace{-0.5cm}
\section{Introduction}

The Milky Way (MW) formed over billions of years through a series of accretion and merger events, whose imprints remain in the form of stellar streams and substructures \citep[e.g.,][]{2004ASPC..317..256M,2018MNRAS.478..611B,2018Natur.563...85H,2019A&A...631L...9K,2019ApJ...872..152I,2020ApJ...901...48N,2022ApJ...926..107M}. These streams, originating from the tidal disruption of dwarf galaxies and globular clusters or from perturbations of the disk, preserve valuable orbital and chemical information. Since the orbital properties of stellar streams and substructures are effective tracers of a galaxy’s formation history and gravitational potential \citep{2010ApJ...714..229L}, their accurate measurement is essential for understanding the origin and evolution of these features, as well as the broader assembly of the MW.

Recently, \citet{2023MNRAS.520.5225M} published the {\em Galstreams} library, a uniform compilation of the orbital parameters for nearly hundred known stellar streams. This work also assessed uncertainties in individual stream parameters, providing guidance for future improvements. Complementary efforts based on Gaia EDR3 and ground-based spectroscopy have identified additional substructures. Using a hierarchical clustering method with integrals of motion (\emph{E}, $L_z$, $L_\perp$), \citet{2022A&A...665A..57L} unveiled 6 main groups or substructures, with further population properties discussed in \citet{2022A&A...665A..58R}. Together, these studies reinforce the role of stellar streams and substructures as powerful tracers of the MW’s assembly history.

Within this framework, the Gaia–Sausage–Enceladus (GSE) has been identified as one of the most prominent accreted components, originally interpreted as the remnant of a single massive merger that shaped the inner halo and thick disk \citep{2018Natur.563...85H,2018MNRAS.478..611B}. However, recent studies \citep{2022ApJ...932L..16D,2023ApJ...944..169D} reveal kinematic and chemical diversity within GSE inconsistent with a single origin, suggesting instead multiple accretion events and distinct substructures such as the Virgo Radial Merger, Cronus, Nereus, and Thamnos. These findings highlight the complex assembly of the MW and emphasize the significance of accretion in shaping globular clusters and dwarf galaxies, which preserve crucial information about the Galaxy’s formative phases and continue to influence the halo’s dynamical and chemical evolution.

\section{Data and Methods}

For the local halo, we adopt the compilation of \citet{2022A&A...665A..57L}, which combines Gaia EDR3 RVS \citep{2021A&A...649A...1G} with complementary radial velocities from LAMOST DR6 \citep{2019RAA....19...75L}, RAVE DR6 \citep{2020AJ....160...82S}, GALAH DR3 \citep{2021MNRAS.506..150B}, and APOGEE DR16 \citep{2020ApJS..249....3A}. This results in a sample of 51,671 stars. Besides, we also use the data from the DESI Early Data Release \citep{2016arXiv161100036D}, adopting the stellar catalog of \citet{2024ApJS..273...19Z}, which provides abundances and kinematics for 136,877 stars after selection.

For the inner halo, we use the LAMOST DR5 catalog of K giant stars \citep{2014ApJ...790..110L}, cross-matched with Gaia DR3. After applying criteria, we obtain a final sample of 8,099 K giants with kinematic information.

We apply the GS$^3$ Hunter method to identify and analyze cluster candidates. Compared to earlier approaches, it integrates Mahalanobis and Euclidean distances for more accurate clustering, employs neural networks for efficiency on high-dimensional data, and can detect both cold and hot stellar streams. Detailed procedures are referred to \citet{2024ApJ...974..219W, 2025ApJ}.

\section{Results}
\subsection{Local Halo Results}

After applying our method to the local halo data sets, we detect 38 clusters near the Sun, of which 21 satisfy our candidate criterion (data fraction $>$ 95\%). Thirteen of these are associated with known substructures, including GSE \citep{2018Natur.563...85H, 2018MNRAS.478..611B}, the hot thick disk \citep{2019A&A...632A...4D, 2018Natur.563...85H}, L-RL3 \citep{2023A&A...670L...2D}, Thamnos \citep{2018ApJ...860L..11K}, the Helmi stream \citep{2022A&A...668L..10R}, and ED-1 \citep{2023A&A...670L...2D}, with cluster \#11 matching cluster \#38 from \citet{2022A&A...665A..58R}. The remaining eight clusters are presented as new discoveries.

\begin{figure*}[!h]
  \centering
  \includegraphics[width=0.99\textwidth, trim=0 120 0 120,
  clip]{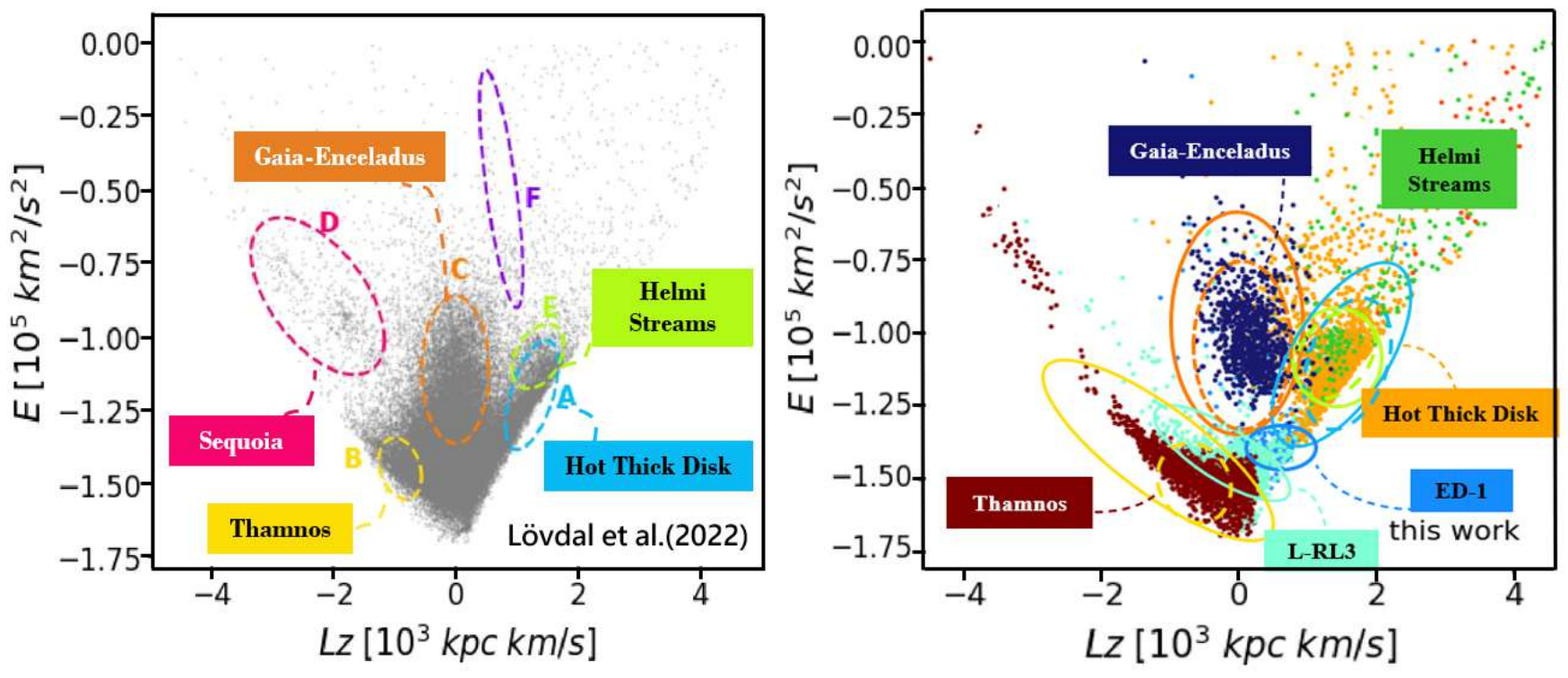}
  \caption{The left panel shows the six main groups in the {\tt\emph E}–{\tt\emph L$_z$} plane from \citet{2022A&A...665A..57L}. The right panel presents our results, with colored labels: solid lines mark our $3\sigma$ substructure regions, while dashed lines show those from \citet{2022A&A...665A..57L}. Our regions are generally more extended, especially for the Hot Thick Disk. Differences reflect sample coverage and methodology, though several structures overlap.}
  \label{comparison}
\end{figure*}

As shown in Figure \ref{comparison}, we recover clustering features in regions consistent with those reported by \citet{2022A&A...665A..57L}, with different colors marking the corresponding substructures. Notably, GSE, Thamnos, and the Helmi stream appear more extended in energy–angular momentum space, likely due to methodological differences. The newly identified clusters and streams provide valuable targets for future studies of their stellar populations and chemical properties.

Recent studies suggest that the Gaia–Sausage–Enceladus (GSE) may comprise multiple components with distinct origins \citep[e.g.,][]{2023ApJ...944..169D}. By this view, we applied GS$^3$ Hunter to the GALAH DR3 sample (using the same selection as \citealt{2023ApJ...944..169D}) and identified 30 clusters. Our results show good agreement with \citet{2023ApJ...944..169D}. Given the complexity of the local halo, different GSE selection strategies inevitably capture substructure mixtures \citep{2022ApJ...932L..16D}.

\begin{figure*}[!h]
  \centering
  \includegraphics[width=0.95\textwidth, trim=0 120 0 120,
  clip]{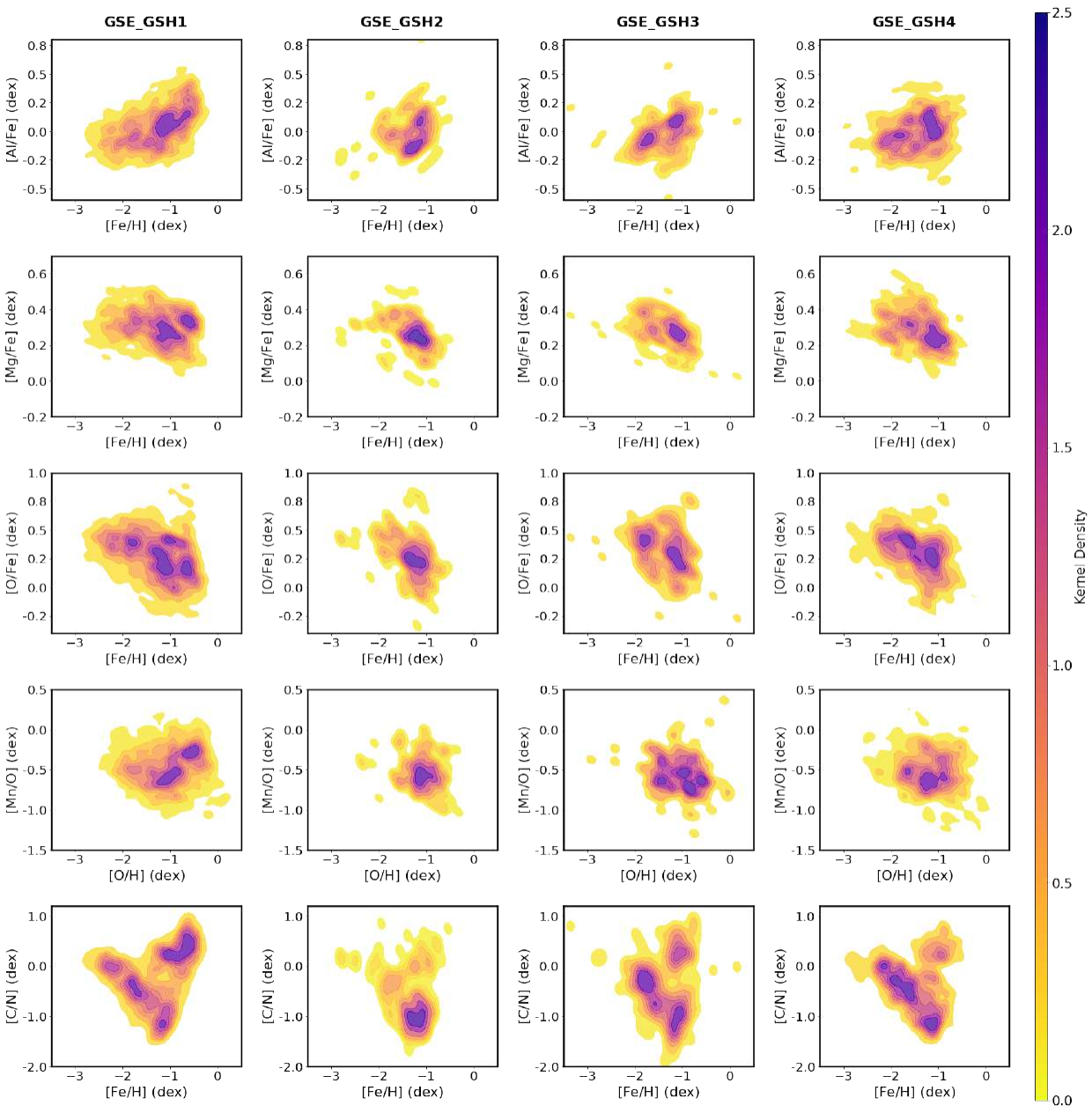}
  \caption{This figure presents the KDE distributions of the four GSE-related structures in chemical abundance space. A colorbar on the right indicates the density levels. In many of the chemical abundance planes, multiple clumps are visible, which may correspond to distinct chemical evolution pathways.}
  \label{DESI(results)_GSE(chemical abundance)}
\end{figure*}

To further constrain and explore the origin of the Gaia–Sausage/Enceladus (GSE) structure, we extend the analysis with DESI EDR data, GS$^3$ Hunter identifies 17 structures. Within the GSE region, we resolve four distinct components, labeled GSE-GSH1 through GSE-GSH4. Kernel density maps in $\alpha$–[Fe/H] chemical space (Figure \ref{DESI(results)_GSE(chemical abundance)}) reveal multiple overdensities within each component, likely tracing separate star-forming episodes or chemically distinct ISM regions.

\subsection{Inner Halo Results and Simulation results}

In the inner halo region, we detect 45 clusters, of which 29 are candidate groups, including 4 linked to Sagittarius \citep{2001ApJ...547L.133I}, 4 to the Virgo Overdensity \citep{2002ApJ...569..245N}, and 2 to the Hercules–Aquila Cloud \citep{2007ApJ...657L..89B}, with six more reported in recent literature. The remaining 13 groups (1891 K-giants) show no clear association with previously known structures, to be explored in future study. We further tested GS$^{3}$ Hunter on the FIRE-2 “Latte” and “ELVIS on FIRE” simulations \citep{2016ApJ...827L..23W, 2021ApJ...920...10P}, which provide rich datasets of Milky Way–like galaxies. Applying our method, we identified 33 groups, including eight true progenitors. GS$^{3}$ Hunter successfully recovers most substructures, demonstrating its robustness.

\section{Summary}

We anticipate that GS$^3$ Hunter will
become a useful tool for the community dedicated to the search for stellar streams and structures in the Milky Way
(MW) and the Local Group, thus helping advance our understanding of the stellar inner and outer halos and the
assembly and tidal stripping history in and around the MW.

\end{document}